\documentclass[onecolumn,pra]{revtex4}
\usepackage{bm}
\usepackage{epsfig}
\usepackage{amsmath}
\begin{document}
\title{Understanding looping kinetics of a long polymer molecule in solution.  Exact solution for delocalized sink model} 

\author{Moumita Ganguly}
\email[]{mouganguly09@gmail.com}
\author{Anirudhha Chakraborty}
\affiliation{School of Basic Sciences, Indian Institute of Technology Mandi, Kamand, Himachal-Pradesh 175005, India.}
\begin{abstract}
 \noindent 
The fundamental understanding of loop formation of long polymer chains in solution has been an important thread of research for several theoretical and experimental studies. Loop formations are important  phenomenological parameters in many important biological processes. Here we give a general method for finding an exact analytical solution for the occurrence of looping of a long polymer chains in solution modeled by using a Smoluchowski-like equation with a delocalized sink. The average rate constant for the delocalized sink is explicitly expressed in terms of the corresponding rate constants for localized sinks with different initial conditions. Simple analytical expressions are provided for average rate constant. 
\end{abstract}
\maketitle
\noindent 
It is well known that loop formation is an important step in several biological processes such as control of gene expression \cite{Rippe, Hippel}, DNA replication \cite{Sun}, protein folding \cite{Eton} and RNA folding \cite{Thiru}. With the progress of single molecule spectroscopic methods, the kinetics of loop formation has regenerated attention of both experimentalists \cite{Winnik, Haung, Hudgind} and theoreticians \cite{Wilemski, Doi, Szabo, Pastor, Portman, Sokolov, Toan},. The advancement in single molecule spectroscopy have made it possible to look into the fluctuations necessary for loop formation at the single molecule level \cite{Widom, Hyeon}.  Loop formation in polymers are actually very complex and exact analytical solution for the dynamics is not possible. All the theories of loop formation dynamics are in general approximate \cite{Doi, Pastor}. In this paper, we give a general method for finding an exact analytical solution for the problem of looping of a long chain polymer in solution. We start with the most simplest one dimensional description of the end-to-end distance of the polymer. The probability distribution $P(x,t)$ of end-to-end distance of a long open chain polymer at time $t$  is given by \cite{Schulten,Moumita}, 
\begin{equation}
\frac{\partial P(x,t)}{\partial t} = \left(\frac{4 N b^2}{\tau_{R}} \frac{\partial^2}{\partial x^2} + \frac{2}{\tau_{R}} \frac{\partial}{\partial x} x -k_{s} - S(x) \right) P(x,t),
\end{equation}
where $x$ denotes end-to-end distance. $\tau_{R}$ is the relaxation time to convert from one to another configuration, length of the polymer is given by $N$ and `$b$' denotes the bond length. The term $k_s$  denotes the rate constant of all other chemical reactions (involving at least one of the end group) apart from the end-to-end loop formation. The occurrence of the looping reaction is given by adding the sink  $S(x)$ term in the R.H.S. of the above equation. The loop formation would occur approximately in the vicinity of the point $x=0$. Therefore it is interesting to analyze a model, where looping occurs at a particular end-to-end distance modelled by representing sink function $S(x)$ by a  Dirac Delta function \cite{Moumita}. Although delocalized sink function provides a more realistic description of the process of looping. The purpose of the current work is to provide an exact solution for rate constants for a general delocalized sink function. It will be further seen that the rate constant for the generalised sink problem in the absence of all other chemical reactions (involving at least one of the end group) apart from the end-to-end loop formation can be evaluated in terms of the corresponding rate constants for localized sinks, with suitable sink positions and initial conditions. To solve Eq. (1), we first use the following transformation
\begin{equation}
P(x,t)=F(x,t) exp (- k_{s} t)
\end{equation}
and obtain the following simplified equation
\begin{equation}
\frac{\partial F}{\partial t}= \frac{4 N b^2}{\tau_{R}} \frac{\partial^2 F}{\partial x^2}+\frac{2}{\tau_{R}}\frac {\partial}{\partial x}\left(x F\right) -S (x)F
\end{equation} 
The solution of Eq. (3) can be written as
\begin{equation}
F(x,t|x_0,0)= F_0 (x,t|x_0,0) -\int^\infty_{-\infty} dx' \int^t_0 dt' F_0(x,t-t'|x',0)\times S(x') F(x',t'|x_0,0),
\end{equation}
where the function $F_0(x,t|x_0,0)$ is the solution in the absence of any sink term, {\it i.e.,} $S(x)=0$. Both $F(x,t|x_0,0)$ and $F_0 (x,t|x_0,0)$ corresponds to the same initial condition, which we consider here to be a Dirac delta function given by the equation
\begin{equation}
F(x,t)=F_0(x,t)=\delta(x-x_0)\;\;\;\; {\text at\; t=0}
\end{equation}
In the following we use the method of Szabo, Lamm, and Weiss \cite{Szabo} where an arbitrary sink function $S(x)$ can be expressed as $S(x) = \int^\infty_{-\infty} dx' S(x')\delta(x-x')$ and the integral can be discretized as shown below
\begin{equation}
S(x)=\sum^N_{i=1} k_i\delta(x-x_i),
\end{equation} 
where $k_i[=\omega_i S(x_i)]$ denotes the sink strengths, with $\omega_i$ depending  on the scheme of discretization. So now Eq.(4) becomes
\begin{equation}
F(x,t|x_0,0)= F_0(x,t|x_0,0) -\sum^N_{i=1} k_i \int^t_0 dt' F_0 (x,t-t'|x_i,0)\times F(x_i,t'|x_0,0).
\end{equation}
Taking appropriate Laplace transform of the above equation, we obtain
\begin{equation}
\tilde{F}(x,k_s+s|x_0,0)= \tilde{F}_0(x,k_s+s|x_0,0) -\sum^N_{j=1} k_j \tilde{F}_0 (x,k_s+s|x_j,0) \times \tilde{F} (x_j,k_s+s|x_0,0)
\end{equation}
where 
\begin{eqnarray}
\tilde{F}(x,k_s+s|x_0,0)= \int^\infty_0 dt\;\;exp[-(k_s+s)t]F(x,t|x_0,0) \\ \nonumber
\tilde{F_0}(x,k_s+s|x_0,0)= \int^\infty_0 dt\;\;exp[-(k_s+s)t]F_0(x,t|x_0,0)\nonumber
\end{eqnarray}
Considering Eq. (7) at the discrete points $x=x_1,x_2,....., x_N, $ we obtain a set of linear equations, which can be written as
\begin{equation}
{\hat A}{\hat P}={\hat Q},    
\end{equation}
where the elements of the matrices $ {\hat A} \rightarrow \{ a_{ij}\}, \; {\hat P} \rightarrow \{ p_{ij}\}$ and $ {\hat Q} \rightarrow \{q_{ij}\}$ are given by 
\begin{eqnarray}
a_{ij}=k_j\tilde{F}_0 (x_i,k_s+s|x_j,0)+\delta_{ij},\\ \nonumber
P_i= \tilde{F}(x_i,k_s+s|x_0,0),\\ \nonumber
q_i = \tilde{F}_0 (x_i,k_s+s|x_0,0).\nonumber
\end{eqnarray}
One can solve the matrix equation {\it i.e.,} Eq. (10) easily and obtain $\tilde{P}(x_i,k_s+s|x_0,0)$ for all $x_i$. Here the quantity of interest is the survival probability $F(t)$ of the open chain polymer, which is defined as
\begin{equation}
F(t) = \int^\infty_{-\infty} P (x,t)dx = exp (-k_{s}t)\left[ 1-\sum^N_{i=1} k_i \int^t_0 F(x_i,t'|x_0,0)dt'\right],
\end{equation} 
so that one can define the average rate constant $k_I$ as \cite{Moumita}
\begin{equation}
k_{I}=\int_{0}^\infty F(t) dt
\end{equation}
and also a  long-time rate constant $k_L$ as\cite{Moumita}
\begin{equation}
K_L =- \lim_{t\rightarrow \infty}(d/dt) In F(t).
\end{equation}
So one can easily show  $k^{-1}_I = \tilde{F}(0)$
and $k_L$ is the negative pole of $\tilde{F}(s)$ closest to the origin.
The average rate constant $k_I$  is thus given by 
\begin{equation}
k^{-1}_I = \lim_{s \rightarrow 0} (k_s+s)^{-1} \left[1- \sum^N_{i=1} k_i \tilde{F} (x_i, k_s+s|x_0,0)\right]
\end{equation} 
where $\tilde{F}(x_i,k_s+s|x_0,0)$ is to be obtained by solving Eq. (10), Which is straightforward if the Laplace transformed quantities $\tilde{F}_0 (x_i,k_s+s|x_j,0)$ and $\tilde{F}_0 (x_i,k_s+s|x_0,0)$ appearing in the matrices ${\hat A}$ and ${\hat Q}$, respectively, can be evaluated analytically. The easiest way to solve Eq. (10) is by Cramer's method, and the solution for $P_j \rightarrow \tilde{F} (x_j,k_s+s|x_0,0) $ is given by 
\begin{equation}
P_j= det {\hat A}^{(j)} / det {\hat A},
\end{equation}
where det ${\hat A}$ represents the determinant of matrix ${\hat A}$ and ${\hat A}^{(j)}$ is a matrix obtained by replacing the j-th column of the matrix ${\hat A}$ by the column vector ${\hat Q}$. Substituting this solution into Eq. (14) one has the result
\begin{equation}
 k^{-1}_I= \lim_ {s\rightarrow 0}\left[ det {\hat A} - \sum^N_{j=1} k_j det {\hat A}^{(j)}\right]/\{(k_s+s)det {\hat A}\}. 
\end{equation}
Using simple algebraic manipulations, it is straightforward to show that the numerator of Eq.(17) remains unchanged if in all the elements of the matrices,  the quantities $\tilde{F}_0(x_i,k_r+s|x_j,0)$ and $\tilde{F}_0(x_i,k_r+s|x_0,0)$ are replaced, respectively, by $\triangle\tilde{F}_0(x_i,k_s+s|x_j,0)$ and $\triangle\tilde{F}_0(x_i,k_s+s|x_0,0),$ defined below.
\begin{equation}
\triangle {\tilde F}_{0}(x,t|x',0)= F_0 (x,t|x_0,0) -\int^\infty_0  dt exp[ -(k_s+s)t] \left( F_{0}(x,t|x',0) - F_0(x, t= \infty) \right),
\end{equation}
In the following we will use the notation that denotes $F^{st}_0=F_0(x, t= \infty)$.
Denoting the modified ${\hat A}$ matrix as matrix ${\hat B}$ with elements $b_{ij}\rightarrow k_j\triangle
\tilde{F}_0 (x_i,k_s+s|x_j,0)+ \delta_{ij}$ , the numerator of Eq. (17) becomes $(det {\hat B}- 
\sum^N_j=1 k_j det {\hat B} ^{(j)}$, where in the matrix ${\hat B}^{(j)}$ the j-th column of matrix ${\hat B}$
has been replaced by the column vector $Q' \rightarrow \{q'_i\}$ with $q'_i = \
\triangle\tilde{F}_0 (x_i,k_s+s|x_0,0).$ \\

Form Eq. (18), one has $\triangle\tilde{F}_0(x_i,k_r+s|x_j,0)=  \tilde{P}_0(x_i,k_s+s|x_j,0)- \
(k_s+s)^{-1} F^{st}_0 (x_j),$ and hence the denominator of Eq. (17) can be rewritten as
\begin{equation}
(k_s+s)det {\hat A}= \left[(k_s+s)det {\hat B}+ \sum^N_{j=1} k_j det{\hat B}^{(j)'}\right]
\end{equation}

where the matrix ${\hat B}^{(j)'}$ is obtained from matrix ${\hat B}$ by replacing the elements $b_{ij}$ of its jth column by the stationary values $F^{st}_0 (x_i)$ for all the rows, {\it i.e.,} i=1,...., N. Thus, on taking the limit $s \rightarrow 0$, the 
final expression for the rate constant $k_I$ is given by \cite{Swapan}

\begin{equation}
 k^{-1}_I = \left[ det {\hat B} - \sum^N_{j=1} k_j det{\hat B}^{(j)}\right]/ \left[ k_s det {\hat B} 
 +\sum^N_{j=1} k_j det {\hat B}^{(j)'}\right],
\end{equation}

 So we have the rate constant for a delocalized sink, for the special case of a localized sink at a point $x_1$, i.e., S(x) = $k_1 \delta (x-x_1)$, it can be expressed in a simple form as below
\begin{equation}
k_I=\{k_1 F^{st}_0 (x_1) +k_s[1+k_1 \triangle \tilde{F}_0(x_1,k_s|x_1,0)]\}/ \{1+k_1[\triangle \tilde{F}_0(x_1,k_s|x_1,0)-\triangle \tilde{F}_0(x_1,k_s|x_0,0)]\}
\end{equation}
For purely looping problem (no other chemical reactions involving end groups), {\it i.e.,} $k_s=0$, the rate constant $K_I$ for the general delocalized sink is given by Eq.(20) can be re-expressed in terms of the rate constant of Eq.(21). The quantities that will appear in the final expression are denoted here as $ k_0(x_i,x')$ which is the rate constant (in case of $k_s=0)$ for single Dirac $\delta$-function sink  $S(x)=k_i\delta(x-x_i)$ if the end-to-end distance initially is $x=x'$ and is given by
\begin{equation}
k^{-1}_0 (x_i,x') = [ k_i F^{st}_0 (x_i)]^{-1} +kd^{-1}(x_i,x'),
\end{equation}
where $k_d(x_i,x')$ represents the overall rate constant in the limit $k_i \rightarrow \infty)$, defined as 
\begin{equation}    
k^{-1}_d (x_i,x')= [1/F^{st}_0(x_i)]\times \int^\infty_0 dt [F_0 (x_i,t|x_i,0)-F_0 (x_i,t|x',0)].
\end{equation} 
After doing little bit of algebra one finally obtains the general rate constant given by
\begin{equation}
k_I^{-1}=[- det {\hat C}+ \sum^{N}_{j=1} det{\hat C}^{(j)}]/ [\sum^{N}_{j=1} det{\hat C}^{(j')}],
\end{equation}
where the elements of the matrix ${\hat C} \rightarrow \{c_{ij}\}$ are given by \\

\begin{equation}
c_{ij}= \{
\begin{array}{cc}
k^{-1}_0(x_i,x_j)& for \;\;i \neq j\\
0 & for\;\; i=j.\\
\end{array}
\end{equation}

The matrix ${\hat C}^{(j)}$ is obtained from the matrix ${\hat C}$ by replacing only its j-th column by the column vector $D=\{d_i\}$, where
\begin{equation}
d_i= k^{-1}_0 (x_i,x_0).
\end{equation}
Similarly, the matrix $\hat{C}^{(j)}$ is same as matrix ${\hat C}$, but for the elements of the j-th column all of which are replaced by unity.
So for $k_s= 0$, the rate constant $k_I$  for looping of a long polymer modelled using a sink of arbitrary shape, described by a set of $N$ localized sinks, can be obtained if the corresponding localized sink rate constant $ k_0 (x_i,x')$ defined in Eq. (22) can be evaluated for arbitrary values of initial end-to-end distance $(x')$, sink position $(x_i)$, and strength $(k_i)$. So in the absence of any sink term
\begin{equation}
F_0(x,t|x',0)=[(\gamma/2\pi)(1-e^{-2\gamma D t})]^{1/2} \times \; exp \left[ \frac{-(\gamma/2)(x - x' e^{-\gamma t})^2}{(1-e^{-2\gamma D t})}\right],
\end{equation}
where $D= \frac{4 N b^2}{\tau_{R}}$ and $\gamma = \frac{1}{2 N b^2}$ and therefore
\begin{equation}
F^{st}_0 (x_i)= (\gamma/2\pi)^{1/2} \; exp [- (\gamma/2) {x_i}^2]
\end{equation}
So $k_d$ can be expressed as
\begin{eqnarray}
k^{-1}_d (x_i,x')=(\gamma D)^{-1} \int^\infty_0 dy [ 1- exp (- 2y)]^{-1/2} [ exp[\gamma x^2_i e ^{-y}/(1+e^{-y})]\\ \nonumber
- exp (\gamma \{ x_i,x' e^{-y}- (e^{-2y}/2) [x^2_i +(x')^2]\}/(1-e^{-2y}))]. \nonumber
\end{eqnarray}
Although in general, it might be difficult to evaluate the above integration analytically, simplified expression can be derived in the case where the sink position is at the origin $(x_i=0)$,  the expression for $k_d(x_i,x')$ simplifies to \cite{Swapan}
\begin{equation}
k^{-1}_d (0,x') = (\pi/2\gamma)^{1/2} D^{-1} \times \int^{|x'|}_0 dx \; exp (\gamma x^2/2)\times \{1-erf[(\gamma/2)^{1/2}x]\},
\end{equation}
The case of initial end-to-end distance of the poplymer is zero ( {\it i.e.,}x'=0) we get \cite{Swapan}
\begin{equation}
k^{-1}_d (x_i,0) = (\pi/2\gamma)^{1/2} D^{-1} \times \int^{|x_i|}_0 dx \; exp (\gamma x^2/2)\times \{1+erf[(\gamma/2)^{1/2}x]\}.
\end{equation}
One can derive a generalized expression \cite{Sebastian}
\begin{equation}
k^{-1}_d (x_i,x') = (\pi/2\gamma)^{1/2} D^{-1} \times \int^{|x_i|}_{x'} dx \; exp (\gamma x^2/2)\times \{1+erf[(\gamma/2)^{1/2}x]\}.
\end{equation}
So $k^{-1}_d (x_i,x')$ can be expressed as a linear combination of $k^{-1}_d (0,x')$ and $k^{-1}_d (x_i,0)$ and this helps a lot in the calculation using a sink of arbitrary shape. So far we have considered only the initial condition $F(x,0)=\delta(x-x_0)$. Now we will consider the case where initial condition has the following distribution, {\it i.e.,} $F(x,0)= F^{st}_0(x)$. In this case, the rate constant $k_I$ for the  general sink is again given by Eqs. (24) and (25),but the expression for $d_i$ given by Eq.(26) is to be replaced by the following expression
\begin{equation}
d_i=[k_iF^{st}_0 (x_i)]^{-1}+[k^{st}_d (x_i)]^{-1},
\end{equation}
where
\begin{equation}
k^{st}_d(x_i)=(\gamma D)[(\gamma/ 2 \pi)x^2_i]^{1/2} \; exp [-(\gamma/2){x_i}^2]/\phi(x_i),
\end{equation}
with 
\begin{eqnarray}
\phi(x_i)=erf\{[(\gamma/2)x^2_i]^{1/2}\}+[(\gamma/2)x^2_i]^{1/2} exp [-(\gamma/2)x^2_i]\\\nonumber
\times \left[- 2+In 2 +(\gamma/2)x^2_i \sum^\infty_{k=0} \sum^\infty_{m=0} (-1)^k[(\gamma/2)x^2_i]^{k+m}/[(k!m!(m+\frac{1}{2})(k+m+1)]\right]\nonumber
\end{eqnarray}
The cardinal result of this work is Eq. (24), where the overall rate constant for looping of a long chain polymer for a generalized sink is expressed in terms of localized sink rate constants with different values for the sink position and initial positions. The simple analytical expression for average rate constant for a generalised sink function is derived.\newline 
\noindent  One of the author (M.G.) would like to thank IIT Mandi for HTRA fellowship and the other author thanks IIT Mandi for providing CPDA grant.

\end{document}